\documentclass[oneside]{amsart}
\usepackage[T1]{fontenc}
\usepackage{geometry}
\makeatletter

\newcommand{\LyX}{L\kern-.1667em\lower.25em\hbox{Y}\kern-.125emX\spacefactor1000}

\makeatother

\begin{document}

\title{Permutation orbifolds}

\author{P. Bantay\\
Inst. for Theor. Phys.\\
Rolland Eötvös Univ., Budapest}

\maketitle
\begin{abstract}
A general theory of permutation orbifolds is developed for arbitrary twist groups.
Explicit expressions for the number of primaries, the partition function, the
genus one characters, the matrix elements of modular transformations and for
fusion rule coefficients are presented, together with the relevant mathematical
concepts, such as \( \Lambda  \)-matrices and twisted dimensions. The arithmetic
restrictions implied by the theory for the allowed modular representations in
CFT are discussed. The simplest nonabelian example with twist group \( S_{3} \)
is described to illustrate the general theory.
\end{abstract}

\section{Introduction}

While the notion of permutation orbifold has been introduced in \cite{KS},
the first instance of a permutation orbifold computation is the work \cite{FHPV},
where a \( \mathbb {Z}_{2} \) permutation orbifold of the \( E_{8}\times E_{8} \)
heterotic string has been constructed for the study of rank reduction in orbifold
models. Following \cite{KS}, a couple of papers investigated permutation orbifolds,
mostly through considerations related to modular invariance \cite{KSF}. Permutation
orbifold techniques have been applied in \cite{DV3} in the description of second
quantized strings. To our knowledge, the first systematic study of permutation
orbifolds appeared in \cite{BHS}, where the primary content, the genus one
characters and their modular transformations, as well as the fusion rules have
been dealt with for permutation orbifolds with cyclic twist group. These results
have been generalized to arbitrary permutation groups in \cite{PO}, which also
laid the foundations of a geometric interpretation of the results through the
theory of covering surfaces. The mathematics of the construction have been clarified
in \cite{BDM} in the framework of Vertex Operator Algebras. Recently, permutation
orbifolds have been investigated in relation to the orbifold Virasoro master
equation \cite{ovme}.

The present paper is an extension of \cite{PO} : after recalling the basic
results of that paper, it deals with such questions as the matrix elements of
arbitrary modular transformations and the structure of the fusion rules, and
gives an account of the relevant mathematical concepts, e.g. \( \Lambda  \)-matrices
and twisted dimensions. Relevant information about permutation actions have
been gathered in Appendix 2. Many important aspects of permutation orbifolds,
like simple currents, the relation to 3D topological field theories, boundary
conditions and unoriented surfaces, have been left for future publications in
order to maintain the logical coherence of the paper. The structure of the simplest
nonabelian permutation orbifold based on the symmetric group on 3 letters is
presented in Appendix 3 to illustrate the results of the main text.

\section{Primaries and their characters}

Let us first recall the most important results of \cite{PO}. We consider a
permutation group \( \Omega  \) of degree \( n \), and a rational CFT \( \mathcal{C} \).
The \( n \)-fold tensor power of \( \mathcal{C} \) admits the permutations
in \( \Omega  \) as symmetries, consequently one may orbifoldize this tensor
power by the twist group \( \Omega  \). We call the resulting theory the \( \Omega  \)
permutation orbifold of \( \mathcal{C} \), and denote it by \( \mathcal{C}\wr \Omega  \).
Note that, properly speaking, it is an orbifold of the tensor power and not
that of \( \mathcal{C} \), but as we shall see one needs the knowledge of the
latter to determine its properties. Because orbifoldization does not alter the
central charge, that of \( \mathcal{C}\wr \Omega  \) is just \( n \) times
the central charge of \( \mathcal{C} \). The wreath product notation for permutation
orbifolds stems from the following basic fact : if \( \Omega  \)\( _{1} \)
and \( \Omega  \)\( _{2} \) are two permutation groups, then the \( \Omega  \)\( _{2} \)
permutation orbifold of \( \mathcal{C}\wr \Omega _{1} \) is the same as the
\( \Omega _{1}\wr \Omega _{2} \) orbifold of \( \mathcal{C} \), i.e.

\begin{equation}
\label{ass}
\left( \mathcal{C}\wr \Omega _{1}\right) \wr \Omega _{2}=\mathcal{C}\wr \left( \Omega _{1}\wr \Omega _{2}\right) 
\end{equation}
where \( \Omega _{1}\wr \Omega _{2} \) denotes the wreath product of the permutation
groups \( \Omega _{1} \) and \( \Omega _{2} \) with the standard ( imprimitive
) action \cite{wreath}. This fundamental result lies at the heart of most of
what follows, e.g. it already enables us to enumerate the primary fields of
the permutation orbifold \( \mathcal{C}\wr \Omega  \) : these are in one-to-one
correspondence with the orbits of \( \Omega  \) on the set of pairs \( \left\langle p,\phi \right\rangle  \),
where \( p \) is an \( n \)-tuple of primaries of \( \mathcal{C} \), i.e.
a map \( p:\left\{ 1,\ldots ,n\right\} \rightarrow \mathcal{I} \) if we denote
by \( \mathcal{I} \) the set of primaries of \( \mathcal{C} \), while \( \phi  \)
is an irreducible character of the double \( \mathcal{D}(\Omega _{p}) \) (c.f.
\cite{DPR} and \cite{LMP}) of the stabilizer \( \Omega _{p}=\left\{ x\in \Omega \, |\, xp=p\right\}  \),
the action of \( \Omega  \) on the map \( p \) being the obvious one, and
the action of \( x\in \Omega  \) on the pair \( \left\langle p,\phi \right\rangle  \)
given simply by
\begin{equation}
\label{omaction}
x\left\langle p,\phi \right\rangle =\left\langle xp,\phi ^{x}\right\rangle 
\end{equation}
where \( \phi ^{x}(y,z)=\phi (y^{x},z^{x}) \) - note that if \( \phi \in \mathcal{D}(\Omega _{p}) \),
then \( \phi ^{x}\in \mathcal{D}(\Omega _{xp}) \), so that the above action
is well-defined. A simple counting argument gives then the number of primaries
of \( \mathcal{C}\wr \Omega  \) :
\begin{equation}
\label{nrprim}
\left| \mathcal{I}^{\Omega }\right| =\frac{1}{|\Omega |}\sum _{(x,y,z)\in \Omega ^{\left\{ 3\right\} }}s^{|\mathcal{O}(x,y,z)|}
\end{equation}
where \( s=\left| \mathcal{I}\right|  \) denotes the number of primaries of
\( \mathcal{C} \), \( \Omega ^{\left\{ k\right\} } \) is the set of commuting
\( k \)-tuples from \( \Omega  \), and \( \mathcal{O}(x,y,z) \) denotes the
set of orbits ( on the set \( \left\{ 1,\ldots ,n\right\}  \) ) of the group
generated by the triple \( x,y,z \). 

Once we know how to label the different primaries, the next task is to determine
their most important characteristics, the most basic ones being the genus one
characters. The key ingredient to do this is to understand the geometric aspect
of orbifoldization, namely its relation to the theory of covering surfaces.
The argument is as follows : should the twist group \( \Omega  \) be trivial,
i.e. should we consider the \( n \)-fold tensor power of \( \mathcal{C} \),
a primary of this trivial permutation orbifold would be simply an \( n \)-tuple
of primaries of \( \mathcal{C} \), and all interesting quantities would factorize
accordingly. That is, we could interpret this trivial permutation orbifold of
\( \mathcal{C} \) over some surface \( \Sigma  \) as if we were actually considering
\( \mathcal{C} \) over a trivial \( n \)-sheeted covering of \( \Sigma  \).
If the twist group \( \Omega  \) is non-trivial, then we should still consider
\( n \)-sheeted coverings of the surface, but this time we have to allow non-trivial
coverings whose monodromy group is a subgroup of \( \Omega  \), and we have
to sum over all such contributions which may be interpreted as instanton corrections.
In case the surface \( \Sigma  \) has no punctures nor boundaries we have to
consider only unramified coverings, but in general ramified coverings have to
be taken into account because of the twisted sectors.

Let's see how the above rather general considerations apply to the computation
of the genus one characters of \( \mathcal{C}\wr \Omega  \). In this case the
surface under consideration is a torus, and the allowed coverings are characterized
by a homomorphism from the fundamental group \( \mathbb {Z}\oplus \mathbb {Z} \)
of the torus into \( \Omega  \), i.e. by a commuting pair \( x,y\in \Omega  \)
corresponding to the images of the generators of the fundamental group. The
resulting covering surface is in general not connected, its connected components
- which are all tori by Riemann-Hurwitz - being in one-to-one correspondence
with the orbits of the image of the homomorphism, i.e. the subgroup generated
by \( x \) and \( y \). If we denote by \( \mathcal{O}(x,y) \) the set of
these orbits, then each \( \xi \in \mathcal{O}(x,y) \) may be characterized
by three integers \( \lambda _{\xi },\mu _{\xi } \) and \( \kappa _{\xi } \)
(cf. Appendix 2) : 

\begin{itemize}
\item \( \lambda _{\xi } \) is the common length of all \( x \) orbits contained
in \( \xi  \)
\item \( \mu _{\xi } \) is the number of the \( x \) orbits in \( \xi  \), so that
\( |\xi |=\lambda _{\xi }\mu _{\xi } \) gives the length of the orbit \( \xi  \)
\item \( \kappa _{\xi } \) is the unique non-negative integer less than \( \lambda _{\xi } \)
such that \( y^{\mu _{\xi }}x^{-\kappa _{\xi }} \) belongs to the stabilizer
of \( \xi  \) - note that \( xy=yx \) implies that the stabilizers of all
points of \( \xi  \) are equal.
\end{itemize}
If the modular parameter of the original torus is \( \tau  \), then the modular
parameter \( \tau _{\xi } \) of the covering torus corresponding to the orbit
\( \xi  \) may be expressed as 
\begin{equation}
\label{covtor}
\tau _{\xi }=\frac{\mu _{\xi }\tau +\kappa _{\xi }}{\lambda _{\xi }}
\end{equation}
 Based on the above, the genus one character of the primary \( \left\langle p,\phi \right\rangle  \)
of \( \mathcal{C}\wr \Omega  \) reads
\begin{equation}
\label{char1}
\chi _{\left\langle p,\phi \right\rangle }(\tau )=\frac{1}{|\Omega _{p}|}\sum _{(x,y)\in \Omega _{p}^{\left\{ 2\right\} }}\overline{\phi }(x,y)\prod _{\xi \in \mathcal{O}(x,y)}\omega _{p_{\xi }}^{-\kappa _{\xi }/\lambda _{\xi }}\chi _{p_{\xi }}(\tau _{\xi })
\end{equation}
In this last formula, \( p_{\xi } \) denotes the value of \( p \) on the orbit
\( \xi  \) - this is well-defined since \( x,y \) stabilize \( p \), consequently
the latter is constant on the orbits belonging to \( \mathcal{O}(x,y) \) -,
while for a primary \( p \) of \( \mathcal{C} \) 
\begin{equation}
\label{omdef}
\omega _{p}=\exp \left( 2\pi i(\Delta _{p}-\frac{c}{24})\right) 
\end{equation}
denotes its exponentiated conformal weight, i.e. the eigenvalue of the Dehn-twist
\( T:\tau \mapsto \tau +1 \). 

Similar covering surface considerations give at once the expression of the genus
one partition function \( Z_{\Omega } \) of \( \mathcal{C}\wr \Omega  \) in
terms of that of \( \mathcal{C} \) :
\begin{equation}
\label{partfun}
Z^{\Omega }(\tau )=\frac{1}{|\Omega |}\sum _{(x,y)\in \Omega ^{\left\{ 2\right\} }}\prod _{\xi \in \mathcal{O}(x,y)}Z(\tau _{\xi })
\end{equation}

One may also express the higher genus partition functions of \( \mathcal{C}\wr \Omega  \)
in terms of those of \( \mathcal{C} \) \cite{uni}, but the resulting formulae
are much more complicated, basically because by the Riemann-Hurwitz formula
the genus of a connected covering of a genus \( g>1 \) surface is always bigger
then \( g \) .

\section{Modular representation and fusion rules}

The explicit knowledge of the genus one characters allows us to determine the
matrix elements of the modular transformations. It turns out that these matrix
elements, as well as most of the quantities that we shall be interested in later,
depend on a set of quantities of the original theory \( \mathcal{C} \) that
are most economically described by the \( \Lambda  \)-matrix formalism presented
in Appendix 1. In terms of this formalism, the matrix elements of a modular
transformation \( M:\tau \mapsto \frac{a\tau +b}{c\tau +d} \) between the primaries
of the permutation orbifold \( \mathcal{C}\wr \Omega  \) are given for \( c\neq 0 \)
by 
\[
M_{\left\langle p,\phi \right\rangle }^{\left\langle q,\psi \right\rangle }=\]

\begin{equation}
\label{modmat}
\frac{1}{|\Omega _{p}||\Omega _{q}|}\sum _{z\in \Omega }\sum _{x,y\in \Omega _{p}\cap \Omega _{zq}}\overline{\phi }(x,y)\psi ^{z}(x^{a}y^{c},x^{b}y^{d})\prod _{\xi \in \mathcal{O}(x,y)}\omega _{p_{\xi }}^{\frac{a\mu _{\xi }}{c\lambda _{\xi }}}\Lambda _{p_{\xi }}^{(zq)_{\xi }}\left( \frac{a\mu _{\xi }+c\kappa _{\xi }}{c\lambda _{\xi }}\right) \omega _{(zq)_{\xi }}^{\widetilde{\mu }_{\xi }}
\end{equation}
where 
\[
\widetilde{\mu }_{\xi }=\frac{d(c\lambda _{\xi },a\mu _{\xi }+c\kappa _{\xi })^{2}}{c\lambda _{\xi }\mu _{\xi }}\]
and \( (k,n) \) denotes the greatest common divisor of the integers \( k \)
and \( n \), while for \( c=0 \) they are easily expressed in terms of the
exponentiated conformal weights
\begin{equation}
\label{cweight}
\omega _{\left\langle p,\phi \right\rangle }=\frac{1}{d_{\phi }}\sum _{x\in \Omega _{p}}\phi (x,x)\prod _{\xi \in \mathcal{O}(x)}\omega _{p_{\xi }}^{\frac{1}{|\xi |}}
\end{equation}
where \( d_{\phi }=\sum _{x}\phi (x,1) \), for we have 
\[
\left( T^{k}\right) _{\left\langle p,\phi \right\rangle }^{\left\langle q,\psi \right\rangle }=\delta _{\left\langle p,\phi \right\rangle ,\left\langle q,\psi \right\rangle }\omega _{\left\langle p,\phi \right\rangle }^{k}\]

Specializing Eq. (\ref{modmat}) to the case of the modular transformation \( S:\tau \mapsto \frac{-1}{\tau } \)
, we get the result
\begin{equation}
\label{Smat}
S_{\left\langle p,\phi \right\rangle }^{\left\langle q,\psi \right\rangle }=\frac{1}{|\Omega _{p}||\Omega _{q}|}\sum _{z\in \Omega }\sum _{x,y\in \Omega _{p}\cap \Omega _{zq}}\overline{\phi }(x,y)\overline{\psi }^{z}(y,x)\prod _{\xi \in \mathcal{O}(x,y)}\Lambda _{p_{\xi }}^{(zq)_{\xi }}\left( \frac{\kappa _{\xi }}{\lambda _{\xi }}\right) 
\end{equation}
It is an easy exercise to check that the above matrix is symmetric, conforming
with Verlinde's theorem. 

Once we know the matrix elements of the transformation \( S \), we can insert
them into Verlinde's formula to compute the fusion rules of the theory. To this
end, let's denote by \( \mathcal{V}_{g}\left( p_{1},\ldots ,p_{N}\right)  \)
the space of genus \( g \) holomorphic blocks for the insertion of the primaries
\( p_{1},\ldots ,p_{N} \) . By the generalized Verlinde formula of \cite{MS},
we have
\begin{equation}
\label{hb1}
\dim \mathcal{V}_{g}\left( p_{1},\ldots ,p_{N}\right) =\sum _{q\in \mathcal{I}}S_{0q}^{2-2g}\prod ^{N}_{i=1}\frac{S_{qp_{i}}}{S_{0q}}
\end{equation}
which specializes to the usual Verlinde formula \cite{Ver}
\begin{equation}
\label{verlinde}
N_{pqr}:=\dim \mathcal{V}_{0}\left( p,q,r\right) =\sum _{s\in \mathcal{I}}\frac{S_{ps}S_{qs}S_{rs}}{S_{0s}}
\end{equation}
for \( g=0 \) and \( N=3 \). Inserting the expression Eq.(\ref{Smat}) into
Eq.(\ref{hb1}) and performing various summations, we arrive at
\[
\dim \mathcal{V}_{g}\left( \left\langle p_{1},\phi _{1}\right\rangle ,\ldots ,\left\langle p_{N},\phi _{N}\right\rangle \right) =\]

\begin{equation}
\label{hb2}
\frac{1}{|\Omega |}\sum _{z_{1},\ldots ,z_{N}\in \Omega }\sum _{x\in \bigcap _{i=1}^{N}\Omega _{z_{i}p_{i}}}\sum _{\theta \in \Omega _{g}(x|z_{1}p_{1},\ldots ,z_{N}p_{N})}\prod _{i=1}^{N}\frac{\overline{\phi _{i}}\left( c_{i}^{z_{i}},x^{z_{i}}\right) }{|\Omega _{p_{i}}|}\prod _{\xi \in \mathcal{O}(x|\theta )}\mathcal{D}_{\xi }(z_{1}p_{1},\ldots ,z_{N}p_{N})
\end{equation}
 Let's take a look at the new objects entering this formula. First, \( \Omega _{g}(x|p_{1},\ldots ,p_{N}) \)
denotes the set of \( N+2g \)-tuples \( \left[ a_{1},b_{1},\ldots ,a_{g},b_{g},c_{1},\ldots ,c_{N}\right]  \)
of elements of \( \Omega  \) commuting with \( x \), which satisfy the two
conditions 

\begin{enumerate}
\item \( \prod _{i=1}^{g}[a_{i},b_{i}]\prod _{i=1}^{N}c_{i}=1 \) 
\item \( c_{i}\in \Omega _{p_{i}} \) for each \( i=1,\ldots ,N \).
\end{enumerate}
For a given \( \theta \in \Omega _{g}(x|p_{1},\ldots ,p_{N}) \) we denote by
\( \mathcal{O}(x|\theta ) \) the set of orbits of the subgroup generated by
\( x \) and the elements of \( \theta  \), and for a given \( \xi \in \mathcal{O}(x|\theta ) \)

\begin{equation}
\label{hb3}
\mathcal{D}_{\xi }(p_{1},\ldots ,p_{N})=\sum _{q\in \mathcal{I}}S_{0q}^{(2-2g-N)\mu _{\xi }}\prod _{i=1}^{N}\prod _{\zeta \in \mathcal{O}_{i}^{\xi }}\Lambda _{q}^{p_{i}(\zeta )}\left( \frac{\mu _{\xi }\kappa _{\zeta }}{|\xi |}\right) 
\end{equation}
where \( \mu _{\xi } \) denotes the number of \( x \) orbits contained in
\( \xi  \), \( \mathcal{O}_{i}^{\xi } \) is the set of those orbits of the
group \( <x,c_{i}> \) that are contained in \( \xi  \), \( p_{i}(\zeta ) \)
denotes the value of \( p_{i} \) on the orbit \( \zeta \in \mathcal{O}^{\xi }_{i} \)
- which is well defined because both \( x \) and \( c_{i} \) stabilize \( p_{i} \)
-, and \( \kappa _{\zeta } \) denotes the unique non-negative integer less
than \( \frac{|\xi |}{\mu _{\xi }} \) such that 
\[
x^{\kappa _{\zeta }}c_{i}^{-\frac{|\zeta |\mu _{\xi }}{|\xi |}}\]
lies in the stabilizer of \( \zeta  \). 

To get some familiarity with formula Eq.(\ref{hb2}), let's consider the fusion
rules of primaries \( \widehat{p} \) of the following special form : \( \widehat{p}=\left\langle \overrightarrow{p},\phi _{0}\right\rangle  \),
where \( \overrightarrow{p} \) is the constant map with image \( p\in \mathcal{I} \),
while \( \phi _{0}(x,y)=\delta _{x,1} \). In particular, the vacuum of \( \mathcal{C}\wr \Omega  \)
is just \( \widehat{0} \). For this special kind of primaries Eq.(\ref{hb2})
tells us that 
\begin{equation}
\label{sfusion}
N_{\widehat{p}\widehat{q}\widehat{r}}=\frac{1}{\left| \Omega \right| }\sum _{x\in \Omega }N_{pqr}^{\left| \mathcal{O}\left( x\right) \right| }
\end{equation}
and more generally 
\[
\dim \mathcal{V}_{0}\left( \widehat{p}_{1},\ldots ,\widehat{p}_{N}\right) =\frac{1}{\left| \Omega \right| }\sum _{x\in \Omega }\dim \mathcal{V}_{0}\left( p_{1},\ldots ,p_{N}\right) ^{\left| \mathcal{O}\left( \mathcal{x}\right) \right| }\]
It should be stressed that the primaries of the form \( \widehat{p} \) do not
form a subring of the fusion ring, i.e. the fusion product of such primaries
contains in general primaries that are not of the form \( \widehat{r} \). 

The formula Eq.(\ref{hb2}) is quite complicated, but can be computed explicitly
- and effectively - for any permutation group \( \Omega  \). There is no advantage
to restrict one's attention to the fusion rules - i.e. the case \( g=0 \) and
\( N=3 \) -, for the resulting expression is basically of the same complexity
as Eq.(\ref{hb2}). An important simplification occurs if one observes that
the generalized fusion rules of Eq.(\ref{hb2}) are (multivariate) polynomials
in the quantities \( \mathcal{D}_{\xi }(p_{1},\ldots ,p_{N}) \), which in turn
fall into a class of quantities formed from \( \Lambda  \)-matrix elements
of a very specific structure. We term this class of quantities \emph{twisted
dimensions} because they generalize the ordinary dimension of holomorphic blocks.
Their definition is as follows : let \( g \) be a non-negative integer, and
consider a sequence \( p_{1},\ldots ,p_{N}\in \mathcal{I} \) of primaries together
with a corresponding sequence \( r_{1},\ldots ,r_{N}\in \mathbb {Q} \) of rational
numbers. We define the genus \( g \) twisted dimension of the primaries \( p_{1},\ldots ,p_{N}\in \mathcal{I} \)
with characteristics \( r_{1},\ldots ,r_{N} \) via the Verlinde-like formula

\begin{equation}
\label{twd}
\mathcal{D}_{g}\left( \begin{array}{ccc}
p_{1} & \ldots  & p_{N}\\
r_{1} & \ldots  & r_{N}
\end{array}\right) =\sum _{q\in \mathcal{I}}S_{0q}^{2-2g}\prod _{i=1}^{N}\frac{\Lambda _{qp_{i}}(r_{i})}{S_{0q}}
\end{equation}

For some properties of twisted dimensions, see Appendix 1. At this point we
just note that

\begin{equation}
\label{0char}
\mathcal{D}_{g}\left( \begin{array}{ccc}
p_{1} & \ldots  & p_{N}\\
0 & \ldots  & 0
\end{array}\right) =\dim \mathcal{V}_{g}(p_{1},\ldots ,p_{N})
\end{equation}
explaining our choice of terminology. 

While twisted dimensions for arbitrary characteristics share many properties
with those for which all characteristics are zero, there is an important difference
: the latter are always non-negative integers, being the dimensions of spaces
of holomorphic blocks, while the former may be in principle arbitrary complex
numbers. This does not mean that twisted dimensions do not satisfy arithmetic
conditions, as they are the building blocks of the fusion rules of the orbifold
model, and the latter are non-negative integers. For example, in case of \( \mathbb {Z}_{2} \)
permutation orbifolds, there is a set of fusion rules that take the form 
\begin{equation}
\label{Z2fus}
\frac{1}{2}\mathcal{D}_{0}\left( \begin{array}{cccc}
p & p & q & r\\
0 & 0 & 0 & 0
\end{array}\right) \pm \frac{1}{2}\mathcal{D}_{0}\left( \begin{array}{ccc}
p & q & r\\
0 & \frac{1}{2} & \frac{1}{2}
\end{array}\right) 
\end{equation}
Here \( p,q,r\in \mathcal{I} \) are arbitrary primaries, and it follows from
Eq.(\ref{0char}) that the first term equals \( \frac{1}{2}N_{ppqr} \). Because
Eq.(\ref{Z2fus}) should give non-negative integers, we get the result that

\begin{equation}
\label{Ydef}
Y_{pqr}:=\mathcal{D}_{0}\left( \begin{array}{ccc}
p & q & r\\
0 & \frac{1}{2} & \frac{1}{2}
\end{array}\right) 
\end{equation}
should be an integer of the same parity as \( N_{ppqr} \), with absolute value
not greater than the latter. The above result appears already in \cite{BHS},
where the fusion rules of \( \mathbb {Z}_{2} \) permutation orbifolds had been
first presented explicitly, and conforms the conjecture of \cite{Sag}, where
the \( Y \)-matrix had been introduced. Let us note that in the special case
where \( r \) is the vacuum, the matrix element \( Y_{pq0} \) is related to
the trace of the braiding operator \cite{FS}.

The above reasoning may be pursued for other twist groups, and leads to very
stringent arithmetic conditions on the allowed values of the twisted dimensions.
Unfortunately, the resulting conditions are much more complicated than in the
\( \mathbb {Z}_{2} \) case, and their analysis is much more involved. The integrality
of some twisted dimensions also enter in the analysis of the congruence subgroup
property of RCFTs \cite{CG}. The arithmetic properties of twisted dimensions
have essentially a topological origin, for it turns out that they are the partition
functions of so-called Seifert-manifolds, i.e. closed 3-manifolds admitting
a free circle action, in the 3D topological field theory associated to \( \mathcal{C} \),
but the details of this identification would lead us too far afield.

Finally, we note that while we have made no assumption on the characteristics
in the definition of twisted dimensions, it follows from Eq. (\ref{rhoadd})
of Appendix 2 that all twisted dimensions entering a fusion rule have the property
that the sum of their characteristics is an integer.

\section*{Appendix 1 : \protect\( \Lambda \protect \)-matrices}

Let \( r=\frac{k}{n} \) be a rational number in reduced form, i.e. with \( k \)
and \( n \) coprime and \( n>0 \). There exists integers \( x \) and \( y \)
such that \( kx-ny=1 \), i.e. the matrix \( M=\left( \begin{array}{cc}
k & y\\
n & x
\end{array}\right)  \) belongs to \( SL(2,Z) \). If we denote by \( M_{pq} \) the matrix element
of the modular transformation \( \tau \mapsto \frac{k\tau +y}{n\tau +x} \)
between the primaries \( p \) and \( q \), then we can define 
\begin{equation}
\label{Lambdadef}
\Lambda _{pq}(r)=\exp \left( 2\pi ir\left( \Delta _{p}-\frac{c}{24}\right) \right) M_{pq}\exp \left( 2\pi ir^{*}\left( \Delta _{q}-\frac{c}{24}\right) \right) 
\end{equation}
or symbolically \( \Lambda (r)=T^{r}MT^{r^{*}} \), where \( r^{*}=\frac{x}{n} \).

The first thing to note is that while the numbers \( x \) and \( y \) are
not determined uniquely, \( \Lambda _{pq}(r) \) is nevertheless well-defined,
for if \( \widetilde{x} \) and \( \widetilde{y} \) is another pair of integers
such that \( k\widetilde{x}-n\widetilde{y}=1 \), then \( \widetilde{x}=x+mn \)
and \( \widetilde{y}=y+mk \) for some integer \( m \), and consequently \( \frac{\widetilde{x}}{n}=r+m \),
while 
\[
\left( \begin{array}{cc}
k & \widetilde{y}\\
n & \widetilde{x}
\end{array}\right) =\left( \begin{array}{cc}
k & y\\
n & x
\end{array}\right) T^{m}\]
where \( T=\left( \begin{array}{cc}
1 & 1\\
0 & 1
\end{array}\right)  \) denotes the Dehn-twist, and the extra \( m \)-th powers of \( T \) cancel
each other. A similar argument shows that \( \Lambda _{pq}(r) \) does not depend
on the integer part of \( r \). In case \( r=0 \), we have \( k=0 \) and
\( n=1 \), and obviously \( \Lambda (0) \) equals the \( S \)-matrix. Moreover,
the unitarity of the modular representation implies that \( \Lambda (r) \)
is a unitary matrix as well. Let's summarize the basic properties of \( \Lambda (r) \)
: 

\begin{enumerate}
\item Periodicity
\begin{equation}
\label{Lper}
\Lambda _{pq}(r+1)=\Lambda _{pq}(r)
\end{equation}

\item Initial value
\begin{equation}
\label{L0}
\Lambda _{pq}(0)=S_{pq}
\end{equation}

\item Symmetry
\begin{equation}
\label{Lsym}
\Lambda _{pq}(r^{*})=\Lambda _{qp}(r)
\end{equation}

\item Inversion
\begin{equation}
\label{Linv}
\Lambda _{pq}(1-r)=\overline{\Lambda }_{\overline{p}q}(r)
\end{equation}

\item Functional equation
\begin{equation}
\label{Lfunc}
\Lambda \left( \frac{-1}{r}\right) =T^{\frac{1}{r}}ST^{r}\Lambda (r)T^{\widehat{r}}
\end{equation}
where \( \widehat{r}=r^{*}+\left( \frac{1}{r}\right) ^{*}=\frac{1}{kn} \). 
\end{enumerate}
Note that in the special case \( r=0 \), the symmetry property Eq. (\ref{Lsym})
expresses the symmetry of the modular matrix \( S \), while the inversion property
Eq. (\ref{Linv}) is the expression of the fact that the matrix \( S^{2} \)
is the charge conjugation matrix. As to the functional equation Eq. (\ref{Lfunc}),
it is the modular relation \( (ST)^{3}=S^{2} \) in a disguised form. The periodicity
property Eq. (\ref{Lper}) and the functional equation allows in principle to
express \( \Lambda (r) \) in terms of \( S \) and \( T \), e.g. using a continued
fraction expansion of \( r \). Some simple cases are listed below : 

\begin{enumerate}
\item 
\[
\Lambda \left( \frac{1}{n}\right) =T^{-\frac{1}{n}}S^{-1}T^{-n}ST^{-\frac{1}{n}}\]
 
\item 
\[
\Lambda \left( \frac{k}{nk+1}\right) =T^{-\frac{k}{nk+1}}S^{-1}T^{-n}ST^{k}ST^{\frac{n}{nk+1}}\]
 
\end{enumerate}
In particular, we have 
\[
\Lambda \left( \frac{1}{2}\right) =T^{-\frac{1}{2}}S^{-1}T^{-2}ST^{-\frac{1}{2}}=P\]
where \( P \) is the matrix introduced in \cite{Sag} in the context of open
strings on non-orientable surfaces, and which also appeared in \cite{BHS} in
the description of \( \mathbb {Z}_{2} \) permutation orbifolds.

Let's now take a closer look at twisted dimensions defined through Eq. (\ref{twd}).
It follows from Eq. (\ref{Lper}) that they are periodic ( with period 1 ) in
the characteristics. The inversion property Eq. (\ref{Linv}) leads to 
\begin{equation}
\label{Dinv}
\overline{\mathcal{D}_{g}\left( \begin{array}{ccc}
p_{1} & \ldots  & p_{N}\\
r_{1} & \ldots  & r_{N}
\end{array}\right) }=\mathcal{D}_{g}\left( \begin{array}{ccc}
\overline{p_{1}} & \ldots  & \overline{p_{N}}\\
1-r_{1} & \ldots  & 1-r_{N}
\end{array}\right) 
\end{equation}
while the unitarity of the \( \Lambda  \)-matrix leads to 
\begin{equation}
\label{Dalg}
\mathcal{D}_{g}\left( \begin{array}{ccc}
p_{1} & \ldots  & p_{N}\\
r_{1} & \ldots  & r_{N}
\end{array}\right) =\sum _{q\in \mathcal{I}}\mathcal{D}_{h}\left( \begin{array}{cccc}
p_{1} & \ldots  & p_{l} & q\\
r_{1} & \ldots  & r_{l} & r
\end{array}\right) \mathcal{D}_{g-h}\left( \begin{array}{cccc}
\overline{q} & p_{l+1} & \ldots  & p_{N}\\
-r & r_{l+1} & \ldots  & r_{N}
\end{array}\right) 
\end{equation}
where \( 0\leq h\leq g \) and \( 1\leq l<N \), with \( r \) an arbitrary
rational number, generalizing the usual properties of fusion rule coefficients.

\section*{Appendix 2 : Permutation actions}

In this appendix we collect some useful facts about permutation actions, which
are used here and there in the main text.

Consider a \( k \)-tuple of commuting permutations \( \left\langle x_{1},\ldots ,x_{k}\right\rangle  \)
of the same degree, and an orbit \( \xi \in \mathcal{O}(x_{1},\ldots ,x_{k}) \).
Denoting by \( F_{k} \) the free abelian group of rank \( k \) generated by
\( X_{1},\ldots ,X_{k} \) , the assignment \( X_{i}\mapsto x_{i} \) defines
a transitive permutation action of \( F_{k} \) on the points of \( \xi  \).
The stabilizer of a point under this action (because \( F_{k} \) is abelian
all point stabilizers are equal) is again a free abelian group of rank \( k \)
( and of index \( \left| \xi \right|  \) in \( F_{k} \) ), with generators

\[
Y_{i}=\prod _{j=1}^{k}X_{j}^{s_{ji}}\; \; i=1,\ldots ,k\]
 where \( s \) is an integer matrix. By a well known theorem, such a matrix
might be transformed into a unique Hermite normal form \( h \) which still
represents the same subgroup of \( F_{k} \), but which is upper triangular,
its diagonal elements being non-negative integers whose product equals the index
\( \left| \xi \right|  \) of the subgroup, and whose off-diagonal entries \( h_{ij} \)
in the \( i \)-th row satisfy \( 0\leq h_{ij}<h_{ii} \). In other words, all
such transitive actions are characterized up to equivalence by the matrix \( h \).
In particular, for a pair \( <x,y> \) of commuting permutations the orbits
\( \xi \in \mathcal{O}(x,y) \) are characterized by a matrix 
\[
h=\left( \begin{array}{cc}
\lambda _{\xi } & -\kappa _{\xi }\\
0 & \mu _{\xi }
\end{array}\right) \]
with \( \lambda _{\xi }\mu _{\xi }=|\xi | \) and \( 0\leq \kappa _{\xi }<\lambda _{\xi } \). 

Next, consider a permutation group \( \Omega  \) acting transitively on the
set \( X \), and a central element \( z\in Z(\Omega ) \). Because the cyclic
subgroup generated by \( z \) is normal in \( \Omega  \), its orbits form
a block system, in particular they all have the same length \( \lambda  \)
dividing \( |X| \), and they are permuted among themselves by \( \Omega  \).
For an arbitrary \( x\in \Omega  \) , each orbit \( \xi \in \mathcal{O}(z,x) \)
of the subgroup generated by \( z \) and \( x \) may be characterized by a
triple \( (\lambda ,\mu _{\xi },\kappa _{\xi }) \) - obviously \( z \) and
\( x \) commute by the centrality of the former. Let's define 
\begin{equation}
\label{rhodef}
\rho (z,x)=\sum _{\xi \in \mathcal{O}(z,x)}\frac{\kappa _{\xi }}{\lambda }
\end{equation}

The basic property of \( \rho  \) is that 
\begin{equation}
\label{rhoadd}
\rho (z,xy)=\rho (z,x)+\rho (z,y)\; \; (mod\, 1)
\end{equation}
 The proof goes as follows : if we denote by \( \Sigma  \) the set of \( z \)
orbits - remember that they form a block system - and by \( \mathbb {Z}_{\lambda } \)
the additive group of integers mod \( \lambda  \), then we can write \( X=\mathbb {Z}_{\lambda }\times \Sigma  \)
(an observation that goes back to Galois), with \( z \) acting on a pair \( (i,p)\in \mathbb {Z}_{\lambda }\times \Sigma  \)
as 
\begin{equation}
\label{zaction}
z(i,p)=(i+1,p)
\end{equation}
For \( x\in \Omega  \) there exists \( \pi _{x}\in Sym(\Sigma ) \) and \( \tau _{x}:\Sigma \rightarrow \mathbb {Z}_{\lambda } \)
such that 
\begin{equation}
\label{xaction}
x(i,p)=(i+\tau _{x}(p),\pi _{x}(p))
\end{equation}
 Clearly, these quantities satisfy 
\begin{equation}
\label{piprop}
\pi _{xy}=\pi _{x}\pi _{y}
\end{equation}
 
\begin{equation}
\label{tauprop}
\tau _{xy}=\tau _{x}\circ \pi _{y}+\tau _{y}
\end{equation}
because Eq. (\ref{xaction}) defines a permutation action. Each orbit \( \xi \in \mathcal{O}(z,x) \)
is of the form \( \xi =\mathbb {Z}_{\lambda }\times \xi ^{*} \) for some orbit
\( \xi ^{*} \) of \( \pi _{x} \) on \( \Sigma  \), and this correspondence
is one-to-one. An easy computation shows that 
\begin{equation}
\label{kappatau}
\kappa _{\xi }=\frac{1}{\lambda }\sum _{p\in \xi ^{*}}\tau _{x}(p)
\end{equation}
and consequently 
\begin{equation}
\label{rhotau}
\rho (z,x)=\frac{1}{\lambda }\sum _{p\in \Sigma }\tau _{x}(p)
\end{equation}
This last equation together with Eq. (\ref{tauprop}) implies the assertion.

An important consequence of Eq. (\ref{rhoadd}) is that if the product of the
elements \( x_{1},\ldots ,x_{N}\in \Omega  \) belongs to the commutator subgroup
\( \Omega \prime  \) of \( \Omega  \), i.e. \( \prod _{i=1}^{N}x_{i}\in \Omega \prime  \),
then \( \sum _{i=1}^{N}\rho (z,x_{i}) \) is an integer, that is the sum of
the characteristics in Eq. (\ref{hb3}) is an integer.

The dependence of \( \rho  \) on its first variable is more complicated. We
just quote the following result 
\[
\rho (z^{k},x)=m(k,\lambda )\rho (z,x)\; \; (mod\, 1)\]
where \( m \) is a solution of the congruence 
\[
\frac{km}{(k,\lambda )}\equiv 1\; \; (mod\, \frac{\lambda }{(k,\lambda )})\]

\section*{Appendix 3 : The \protect\( S_{3}\protect \) permutation orbifold}

In this appendix we present the structure of \( S_{3} \) permutation orbifolds,
whose twist group is the symmetric group on 3 letters. The computations have
been performed using the software package GAP \cite{gap}. 

Suppose we are given a RCFT \( \mathcal{C} \) with central charge \( c \),
partition function \( Z\left( \tau \right)  \) and primaries in the set \( \mathcal{I} \).
Because the degree of \( S_{3} \) is 3, the central charge of the permutation
orbifold \( \mathcal{C}\wr S_{3} \) is \( 3c \), and from formula Eq.(\ref{nrprim})
we get for the number of its primaries 
\begin{equation}
\label{s3prim}
\frac{1}{6}\left( s^{3}+21s^{2}+26s\right) 
\end{equation}
where \( s=\left| \mathcal{I}\right|  \) is the number of primaries of \( \mathcal{C} \).
The partition function is 
\begin{eqnarray}
Z^{S_{3}}\left( \tau \right) =\frac{1}{6}Z^{3}\left( \tau \right) +\frac{1}{2}Z\left( \tau \right) \left( Z\left( 2\tau \right) +Z\left( \frac{\tau }{2}\right) +Z\left( \frac{\tau +1}{2}\right) \right) + &  & \\
\frac{1}{3}\left( Z\left( 3\tau \right) +Z\left( \frac{\tau }{3}\right) +Z\left( \frac{\tau +1}{3}\right) +Z\left( \frac{\tau +2}{3}\right) \right)  &  & 
\end{eqnarray}

The genus one character \( \chi _{p}\left( \tau \right)  \) of a primary \( p\in \mathcal{I} \)
of conformal weight \( \Delta _{p} \) has the form 
\[
\chi _{p}\left( \tau \right) =q^{\left( \Delta _{p}-c/24\right) }\widehat{\chi }_{p}\left( \tau \right) \]
where \( q=e^{2\pi i\tau } \), and \( \widehat{\chi }_{p}\left( \tau \right) =\sum _{n=0}a_{n}q^{n} \)
is a holomorphic function of \( q \) with non-negative integer coefficients
\( a_{n} \), which we call the regular part of the character.

To enumerate the primaries of \( \mathcal{C}\wr S_{3} \), one has first to
enumerate the maps from \( \left\{ 1,2,3\right\}  \) to \( \mathcal{I} \),
i.e. triples of primaries from \( \mathcal{I} \). To this end it is convenient
to introduce a (arbitrary) total ordering \( < \) on \( \mathcal{I} \), which
we shall assume tacitly from now on. Triples of primaries are partitioned in
three classes :

\begin{enumerate}
\item \( \left\langle p,q,r\right\rangle  \) with no two components equal : \( p\neq q\neq r \).
The stabilizer of such a triple is the trivial subgroup, and the orbits under
\( S_{3} \) of such triples contain a representative with \( p<q<r \).
\item Triples where two components are equal, but differ from the third one. In this
case the stabilizer is a \( \mathbb {Z}_{2} \) subgroup, and each orbit contains
a representative of the form \( \left\langle p,p,q\right\rangle  \) with \( p\neq q \).
\item Triples \( \left\langle p,p,p\right\rangle  \) with all components equal. The
stabilizer is the full group \( S_{3} \).
\end{enumerate}
Let's begin with the first case. Because the stabilizer is trivial, its double
has only one irrep, so we don't need to introduce any extra label, i.e. a triple
\( \left\langle p,q,r\right\rangle  \) with \( p<q<r \) corresponds to a primary
of \( \mathcal{C}\wr S_{3} \). The following table contains respectively the
conformal weight, the Frobenius-Schur indicator (see \cite{FS}) and the regular
part of the genus one character of the corresponding primary.

\vspace{0.3cm}
{\centering \begin{tabular}{|c|}
\hline 
\\
\( \Delta _{p}+\Delta _{q}+\Delta _{r} \)\\
\hline 
\\
\( \nu _{p}\nu _{q}\nu _{r}+C_{pq}\nu _{r}+C_{qr}\nu _{p}+C_{rp}\nu _{q} \) \\
\hline 
\\
\( \widehat{\chi }_{p}\left( \tau \right) \widehat{\chi }_{q}\left( \tau \right) \widehat{\chi }_{r}\left( \tau \right)  \) \\
\hline 
\end{tabular}\par}
\vspace{0.3cm}

The notation \( C_{pq} \) stands for the matrix elements of the charge conjugation
operator, i.e. \( C_{pq}=\delta _{p,\overline{q}} \) .

Let's now consider the second case, i.e. a triple of the form \( \left\langle p,p,q\right\rangle  \)
with \( p\neq q \). As the stabilizer is \( \mathbb {Z}_{2} \) , its double
has four irreps, i.e. to each such triple there corresponds four different primaries.
We'll label them simply by an integer from \( \left\{ 1,\ldots ,4\right\}  \),
and the following table summarizes their properties.

\vspace{0.3cm}
{\centering \begin{tabular}{|c||c|}
\hline 
&
\\
&
\( 2\Delta _{p}+\Delta _{q} \)\\
\cline{2-2} 
&
\multicolumn{1}{|c|}{}\\
1&
\multicolumn{1}{|c|}{\( \frac{1}{2}\nu _{p}\nu _{q}\left( \nu _{p}+1\right)  \)}\\
\cline{2-2} 
&
\multicolumn{1}{|c|}{}\\
&
\multicolumn{1}{|c|}{\( \frac{1}{2}\widehat{\chi }_{q}\left( \tau \right) \left( \widehat{\chi }_{p}^{2}\left( \tau \right) +\widehat{\chi }_{p}\left( 2\tau \right) \right)  \)}\\
\hline 
&
\\
&
\( 2\Delta _{p}+\Delta _{q} \)\\
\cline{2-2} 
&
\multicolumn{1}{|c|}{}\\
2&
\multicolumn{1}{|c|}{\( \frac{1}{2}\nu _{p}\nu _{q}\left( \nu _{p}+1\right)  \)}\\
\cline{2-2} 
&
\multicolumn{1}{|c|}{}\\
&
\multicolumn{1}{|c|}{\( \frac{1}{2}\widehat{\chi }_{q}\left( \tau \right) \left( \widehat{\chi }_{p}^{2}\left( \tau \right) -\widehat{\chi }_{p}\left( 2\tau \right) \right)  \)}\\
\hline 
&
\\
&
\( \frac{1}{2}\Delta _{p}+\Delta _{q}+\frac{c}{16} \)\\
\cline{2-2} 
&
\multicolumn{1}{|c|}{}\\
3&
\multicolumn{1}{|c|}{\( \nu _{p}\nu _{q} \)}\\
\cline{2-2} 
&
\multicolumn{1}{|c|}{}\\
&
\multicolumn{1}{|c|}{\( \frac{1}{2}\widehat{\chi }_{q}\left( \tau \right) \left( \widehat{\chi }_{p}\left( \frac{\tau }{2}\right) +\widehat{\chi }_{p}\left( \frac{\tau +1}{2}\right) \right)  \)}\\
\hline 
&
\\
&
\( \frac{1}{2}\Delta _{p}+\Delta _{q}+\frac{c}{16}+\frac{1}{2} \)\\
\cline{2-2} 
&
\multicolumn{1}{|c|}{}\\
4&
\multicolumn{1}{|c|}{\( \nu _{p}\nu _{q} \)}\\
\cline{2-2} 
&
\multicolumn{1}{|c|}{}\\
&
\( \frac{1}{2}\widehat{\chi }_{q}\left( \tau \right) \left( \widehat{\chi }_{p}\left( \frac{\tau }{2}\right) -\widehat{\chi }_{p}\left( \frac{\tau +1}{2}\right) \right)  \)\\
\hline 
\end{tabular}\par}
\vspace{0.3cm}

Finally, we have the triples of the form \( \left\langle p,p,p\right\rangle  \)
for which the stabilizer is \( S_{3} \), whose double has eight irreducible
characters which we'll simply label by an integer from \( \left\{ 1,\ldots ,8\right\}  \).
Their characteristics look as follows (\( \epsilon =e^{\frac{2\pi i}{3}} \)
is a third root of unity).

\bigskip{}
{\centering \begin{tabular}{|l||c|}
\hline 
&
\\
&
\multicolumn{1}{|c|}{\( 3\Delta _{p} \)}\\
\cline{2-2} 
&
\\
1&
\multicolumn{1}{|c|}{\( \frac{1}{2}\nu _{p}\left( \nu _{p}+1\right)  \)}\\
\cline{2-2} 
&
\multicolumn{1}{|c|}{}\\
&
\multicolumn{1}{|c|}{\( \frac{1}{6}\left( \widehat{\chi }_{p}^{3}\left( \tau \right) +3\widehat{\chi }_{p}\left( \tau \right) \widehat{\chi }_{p}\left( 2\tau \right) +2\widehat{\chi }_{p}\left( 3\tau \right) \right)  \)}\\
\hline 
&
\\
&
\multicolumn{1}{|c|}{\( 3\Delta _{p} \)}\\
\cline{2-2} 
&
\multicolumn{1}{|c|}{}\\
2&
\multicolumn{1}{|c|}{\( \frac{1}{2}\nu _{p}\left( \nu _{p}+1\right)  \)}\\
\cline{2-2} 
&
\multicolumn{1}{|c|}{}\\
&
\multicolumn{1}{|c|}{\( \frac{1}{6}\left( \widehat{\chi }_{p}^{3}\left( \tau \right) -3\widehat{\chi }_{p}\left( \tau \right) \widehat{\chi }_{p}\left( 2\tau \right) +2\widehat{\chi }_{p}\left( 3\tau \right) \right)  \)}\\
\hline 
&
\\
&
\multicolumn{1}{|c|}{\( 3\Delta _{p} \)}\\
\cline{2-2} 
&
\multicolumn{1}{|c|}{}\\
3&
\multicolumn{1}{|c|}{\( \nu _{p}^{2} \)}\\
\cline{2-2} 
&
\multicolumn{1}{|c|}{}\\
&
\multicolumn{1}{|c|}{\( \frac{1}{3}\left( \widehat{\chi }_{p}^{3}\left( \tau \right) -\widehat{\chi }_{p}\left( 3\tau \right) \right)  \)}\\
\hline 
&
\\
&
\( \frac{3}{2}\Delta _{p}+\frac{c}{16} \)\\
\cline{2-2} 
&
\multicolumn{1}{|c|}{}\\
4&
\multicolumn{1}{|c|}{\( \nu _{p}^{2} \)}\\
\cline{2-2} 
&
\multicolumn{1}{|c|}{}\\
&
\multicolumn{1}{|c|}{\( \frac{1}{2}\widehat{\chi }_{p}\left( \tau \right) \left( \widehat{\chi }_{p}\left( \frac{\tau }{2}\right) +\widehat{\chi }_{p}\left( \frac{\tau +1}{2}\right) \right)  \)}\\
\hline 
&
\\
&
\( \frac{3}{2}\Delta _{p}+\frac{c}{16}+\frac{1}{2} \)\\
\cline{2-2} 
&
\multicolumn{1}{|c|}{}\\
5&
\multicolumn{1}{|c|}{\( \nu _{p}^{2} \)}\\
\cline{2-2} 
&
\multicolumn{1}{|c|}{}\\
&
\multicolumn{1}{|c|}{\( \frac{1}{2}\widehat{\chi }_{p}\left( \tau \right) \left( \widehat{\chi }_{p}\left( \frac{\tau }{2}\right) -\widehat{\chi }_{p}\left( \frac{\tau +1}{2}\right) \right)  \)}\\
\hline 
&
\\
&
\multicolumn{1}{|c|}{\( \frac{1}{3}\Delta _{p}+\frac{c}{9} \)}\\
\cline{2-2} 
&
\multicolumn{1}{|c|}{}\\
6&
\multicolumn{1}{|c|}{\( \nu _{p} \)}\\
\cline{2-2} 
&
\multicolumn{1}{|c|}{}\\
&
\multicolumn{1}{|c|}{\( \frac{1}{3}\left( \widehat{\chi }_{p}\left( \frac{\tau }{3}\right) +\widehat{\chi }_{p}\left( \frac{\tau +1}{3}\right) +\widehat{\chi }_{p}\left( \frac{\tau +2}{3}\right) \right)  \)}\\
\hline 
&
\\
&
\multicolumn{1}{|c|}{\( \frac{1}{3}\Delta _{p}+\frac{c}{9}+\frac{1}{3} \)}\\
\cline{2-2} 
&
\multicolumn{1}{|c|}{}\\
7&
\multicolumn{1}{|c|}{\( \nu _{p} \)}\\
\cline{2-2} 
&
\multicolumn{1}{|c|}{}\\
&
\multicolumn{1}{|c|}{\( \frac{1}{3}\left( \widehat{\chi }_{p}\left( \frac{\tau }{3}\right) +\epsilon \widehat{\chi }_{p}\left( \frac{\tau +1}{3}\right) +\overline{\epsilon }\widehat{\chi }_{p}\left( \frac{\tau +2}{3}\right) \right)  \)}\\
\hline 
&
\\
&
\multicolumn{1}{|c|}{\( \frac{1}{3}\Delta _{p}+\frac{c}{9}+\frac{2}{3} \)}\\
\cline{2-2} 
&
\multicolumn{1}{|c|}{}\\
8&
\multicolumn{1}{|c|}{\( \nu _{p} \)}\\
\cline{2-2} 
&
\multicolumn{1}{|c|}{}\\
&
\( \frac{1}{3}\left( \widehat{\chi }_{p}\left( \frac{\tau }{3}\right) +\overline{\epsilon }\widehat{\chi }_{p}\left( \frac{\tau +1}{3}\right) +\epsilon \widehat{\chi }_{p}\left( \frac{\tau +2}{3}\right) \right)  \)\\
\hline 
\end{tabular}\par}
\vspace{1cm}

Note that we get a total of

\[
\frac{s\left( s-1\right) \left( s-2\right) }{6}+4s\left( s-1\right) +8s\]
primaries, in accordance with Eq.(\ref{s3prim}).
\newpage

\thanks{Acknowledgments : Many thanks to Geoff Mason and Christoph Schweigert for the
helpful correspondence. This work was supported by grant OTKA F19477.}


\begin{thebibliography}{}
\bibitem{KS}A. Klemm and M.G. Schmidt, Phys. Lett. B\textbf{245}, 53 (1990).
\bibitem{FHPV}P. Forgacs, Z. Horvath, L. Palla and P. Vecsernyes, Nucl. Phys. B\textbf{308,}
477 (1988).
\bibitem{KSF}J.Fuchs, A. Klemm and M.G. Schmidt, Ann.Phys. \textbf{214}, 221 (1992).
\bibitem{DV3}R. Dijkgraaf, G. Moore, E. Verlinde and H. Verlinde, Commun. Math. Phys. \textbf{185},
197 (1997).
\bibitem{BHS}L. Borisov, M.B. Halpern and C. Schweigert, Int. J. Mod. Phys. A\textbf{13}
, 125 (1998).
\bibitem{PO}P. Bantay, Phys. Lett. B\textbf{419}, 175 (1998).
\bibitem{BDM}K. Barron, C. Dong and G. Mason, math/9803118.
\bibitem{ovme}J. Evslin, M.B. Halpern and J.E. Wang, hep-th/9904105; J.de Boer, J. Evslin,
M.B. Halpern and J.E. Wang, hep-th/9908187.
\bibitem{wreath}A. Kerber, \emph{Representations of Permutation Groups I-II,} Springer (1971).
\bibitem{DPR}R. Dijkgraaf, V. Pasquier and P. Roche, Nucl. Phys. Proc. Suppl. \textbf{18}B,
60 (1990).
\bibitem{LMP}P. Bantay, Phys. Lett. B\textbf{245}, 477 (1990); Lett. Math. Phys. \textbf{22},
187 (1991).
\bibitem{uni}P.Bantay, hep-th/9808023.
\bibitem{MS}G. Moore and N. Seiberg, Commun. Math. Phys. \textbf{123}, 177 (1989).
\bibitem{Ver}E. Verlinde, Nucl. Phys. B\textbf{300}, 360 (1988).
\bibitem{Sag}G. Pradisi, A. Sagnotti and Ya.S. Stanev, Phys. Lett. B\textbf{381,} 97 (1996).
\bibitem{FS}P. Bantay, Phys. Lett. B\textbf{394}, 87 (1997).
\bibitem{CG}A. Coste and T. Gannon, math.QA/9909080.
\bibitem{gap}The GAP Group, GAP --- Groups, Algorithms, and Programming, Version 4.1; Aachen,
St Andrews, 1999. (http://www-gap.dcs.st-and.ac.uk/\~{}gap).
\end{thebibliography}
\end{document}